%% file: main.tex
\documentclass[sigconf]{acmart}
\copyrightyear{2026}
\acmYear{2026}
\setcopyright{cc}
\setcctype{by}
\acmConference[EASE '26]{International Conference on Evaluation and Assessment in Software Engineering}{June 09--12, 2026}{Glasgow, United Kingdom}
\acmBooktitle{International Conference on Evaluation and Assessment in Software Engineering (EASE '26), June 09--12, 2026, Glasgow, United Kingdom}
\acmDOI{10.1145/3816483.3816514}
\acmISBN{979-8-4007-2348-3/2026/06}

\usepackage{float}
\usepackage{subcaption}
\usepackage{tabularx}
\usepackage{todonotes}
\usepackage{tcolorbox}
\usepackage{amsmath}
\usepackage{balance}
\usepackage{url}
\usepackage{caption}
\usepackage{hyperref}
\usepackage{booktabs} 
\usepackage{natbib}

\title{Efficiency for Experts, Visibility for Newcomers: A Case Study of Label–Code Alignment in Kubernetes}

\author{Matteo Vaccargiu}
\affiliation{
  \institution{University of Cagliari}
  \city{Cagliari}
  \country{Italy}
}
\email{matteo.vaccargiu@unica.it}

\author{Sabrina Aufiero}
\affiliation{
  \institution{University College London}
  \city{London}
  \country{United Kingdom}
}
\email{sabrina.aufiero.22@ucl.ac.uk}

\author{Silvia Bartolucci}
\affiliation{
  \institution{University College London}
  \city{London}
  \country{United Kingdom}
}
\email{s.bartolucci@ucl.ac.uk}

\author{Ronnie de Souza Santos}
\affiliation{
  \institution{University of Calgary}
  \city{Calgary}
  \country{Canada}
}
\email{ronnie.desouzasantos@ucalgary.ca}

\author{Roberto Tonelli}
\affiliation{
  \institution{University of Cagliari}
  \city{Cagliari}
  \country{Italy}
}
\email{roberto.tonelli@unica.it}

\author{Giuseppe Destefanis}
\affiliation{
  \institution{University College London}
  \city{London}
  \country{United Kingdom}
}
\email{g.destefanis@ucl.ac.uk}

\begin{document}

\begin{abstract}
Labels on platforms such as GitHub support triage and coordination, yet little is known about how well they align with code modifications or how such alignment affects collaboration across contributor experience levels. We present a case study of the Kubernetes project, introducing label--diff congruence---the alignment between pull request labels and modified files---and examining its prevalence, stability, behavioral validation, and relationship to collaboration outcomes across contributor tiers. We analyse 18,020 pull requests (2014--2025) with area labels and complete file diffs, validate alignment through analysis of over one million review comments and label corrections, and test associations with time-to-merge and discussion characteristics using quantile regression and negative binomial models stratified by contributor experience. Congruence is prevalent (46.6\% perfect alignment), stable over years, and routinely maintained (9.2\% of PRs corrected during review). It does not predict merge speed but shapes discussion: among core developers (81\% of the sample), higher congruence predicts quieter reviews (18\% fewer participants), whereas among one-time contributors it predicts more engagement (28\% more participants). Label--diff congruence influences how collaboration unfolds during review, supporting efficiency for experienced developers and visibility for newcomers. For projects with similar labeling conventions, monitoring alignment can help detect coordination friction and provide guidance when labels and code diverge.
\end{abstract}


\maketitle

\input{introduction}

\input{relatedWork}

\begin{table}[t]
\centering
\small
\setlength{\tabcolsep}{3pt}
\caption{Dataset overview, integrity checks, and bot filtering. Percentages are computed over available rows.}
\label{tab:dataset_overview}
\begin{tabular}{l r}
\toprule
\textbf{Property} & \textbf{Value} \\
\midrule
\multicolumn{2}{l}{\textit{Pull requests}} \\
Total PR & 83{,}368 \\
PR time span & 2014-06-07 -- 2025-03-09 \\
Unique PR authors (all) & 6{,}651 \\
Unique PR authors (human-only) & 6{,}615 \\
Merged PR & 60{,}894 (73.0\%) \\
PR with any label & 94.0\% \\
PR with area/* label & 37.9\% \\
PR with merge commit SHA & 99.0\% \\
PRs authored by bots (count) & 36 \\
\hline
\multicolumn{2}{l}{\textit{Commits}} \\
Total commits & 130{,}832 \\
Commit time span & 2014-06-06 -- 2025-07-01 \\
Unique commit authors (all) & 3{,}756 \\
Unique commit authors (human-only) & 3{,}523 \\
Commits with file list & 97.6\% \\
Changed files per commit (P25/P50/P75) & 1 / 2 / 4 \\
Commits authored by bots & 22.69\% (29{,}685) \\
\hline
\multicolumn{2}{l}{\textit{Joinability (PR $\rightarrow$ merge commit)}} \\
PR with merge SHA & 82{,}554 \\
PR joined with diff files & 41{,}399 \\
PR with diff via merge commit & 50.1\% \\
\hline
\multicolumn{2}{l}{\textit{Comments}} \\
Total comments & 1{,}795{,}423 \\
\bottomrule
\end{tabular}
\end{table}

\input{dataset}

\input{methodology}
\input{rq1}
\input{rq2}
\input{discussion}

\input{threats}

\input{conclusion}

\balance
\bibliographystyle{ACM-Reference-Format}
\bibliography{biblio}

\end{document}

%% file: introduction.tex
\section{Introduction}
Large-scale software projects require coordination among thousands of contributors working across organizational boundaries, time zones, and technical subsystems~\cite{herbsleb2003empirical}. Contributors decide where to focus their effort, maintainers route work to reviewers, and teams must avoid duplicated tasks across workstreams~\citep{jeong2009improving, anvik2006should}. Without effective coordination, distributed projects risk wasted effort, review delays, and uneven participation, which affects efficiency and the experience of contributors, especially newcomers unfamiliar with project structure~\citep{hunsen2020fulfillment}.

On platforms such as GitHub, labels act as coordination signals. They are lightweight annotations attached to issues and pull requests (PRs) that indicate subsystems, workflow states, or priorities~\citep{cabot2015}. For contributors, labels help identify relevant tasks and filter work aligned with their interests~\citep{gousios2015work}; for maintainers, they support triage and reviewer assignment. Their usefulness depends on whether they reflect the content of underlying changes. A PR labeled \texttt{area/networking} (indicating the networking subsystem) that modifies only storage files creates uncertainty: contributors may hesitate to review it, and maintainers may misroute it. For newcomers, such mismatches are particularly difficult to interpret~\citep{herzig2013s}.

Despite their widespread use, little is known about how well labels correspond to code changes, whether alignment varies with contributor experience, or how label quality affects review. Do developers correct mislabels, or rely on mentions and informal channels? Does accurate labeling make review smoother, or faster? Do labels play the same role for experienced contributors and newcomers?

We conduct a case study of Kubernetes because it combines scale (nearly 26,000 contributors in the repository GitHub, ranging from one-time participants to core maintainers) with structure (about 175 labels linked to Special Interest Groups that provide both technical categorization and responsibility assignment)~\citep{cabot2015,mockus2002two,aufiero2026}. This heterogeneity allows us to examine whether coordination mechanisms affect populations in distinct ways, addressing a key question for inclusive participation in distributed projects. All project activity is publicly available through GitHub~\citep{kalliamvakou2014promises}, enabling systematic analysis of how labels relate to code changes and review interactions.

We introduce \textbf{label--diff congruence}, defined as the correspondence between area labels on a PR and the files modified in its diff~\citep{cataldo2008socio}. We examine three properties: \textbf{prevalence} (how common it is), \textbf{stability} (whether it changes over time), and \textbf{interpretability} (whether high values correspond to recognizable domains). We also validate the construct by testing whether developers notice and maintain alignment in practice. Our analysis is guided by two research questions that emphasize coordination:\\
\noindent
\textbf{RQ1: How prevalent, stable, and interpretable is label--diff congruence in Kubernetes pull requests, and how do author activity patterns affect it?} We first establish that congruence can be measured consistently and that developers attend to it in practice. We analyze its distribution across PRs, test its stability, and assess whether high values correspond to identifiable technical domains. We also examine variation by contributor experience and validate the construct by checking whether developers comment on coordination and correct labels when misalignment occurs, showing that it reflects actual practice rather than a measurement artifact.
\noindent
\textbf{RQ2: How does label--diff congruence relate to collaboration speed and smoothness on Kubernetes pull requests?} If labels function as coordination signals, their quality should affect review interactions. We test whether congruence predicts two outcomes: \textbf{speed} (time-to-merge) and \textbf{smoothness} (discussion volume and participant diversity). Speed captures how quickly consensus is reached on a contribution, while smoothness describes whether reviews involve focused discussion among a few experts or broader exchanges with many participants. We also stratify by contributor experience to assess whether labels serve the same role for different groups. Divergent effects would indicate that metadata quality influences participation unevenly, helping experienced developers while posing challenges for newcomers.

The contributions of this paper are threefold. First, we introduce \emph{label--diff congruence} as a construct of coordination quality in PRs. Second, we analyze 18,020 Kubernetes PRs, showing that congruence is prevalent, stable, and actively maintained. Third, we demonstrate that congruence shapes review dynamics differently: reducing discussion for experienced contributors while increasing visibility for newcomers. The study shows how routine metadata accuracy influences participation in large-scale development.

To promote transparency and reproducibility, we provide \textbf{our complete analysis pipeline and datasets in a replication package} at this [\href{https://figshare.com/s/bffe6189a8037fc37585}{link}]\footnote{\url{https://figshare.com/s/bffe6189a8037fc37585}}.

%% file: relatedWork.tex
\section{Related Work}

\paragraph{Issue Labeling and Coordination in Software Projects}
Labels support coordination in distributed projects. Cabot et al.~\citep{cabot2015} analyzed over three million GitHub projects and found that only 3.25\% use labels, yet labeled issues achieve higher resolution rates (43.51\%) than unlabeled ones (22.53\%). They identified label families reflecting different coordination intents such as priority, component, and workflow. Kim and Lee~\citep{kim2021} examined 14{,}415 projects with 13 million issues, reporting that 90\% of projects use multiple labels and that multi-label issues close faster (131.9 vs.\ 173.6 days). Joselito Jr.\ et al.~\citep{joselito2024} studied 13{,}280 repositories, showing that 73.14\% use issue labeling and that most activity occurs within 100 days of issue opening, with temporal patterns tied to triage, implementation, and validation phases. Kallis et al.~\citep{kallis2019} proposed TicketTagger, using fastText to classify issues into bug, enhancement, and question categories with precision of 78.1--82.2\% and recall of 76.3--87.4\%. Santos et al.~\citep{santos2023tag} examined API-domain label assignment in issue tracking systems, extending label classification to domain-specific categories tied to software components.  These studies focus on label presence, quantity, and automation, whereas our work examines \textit{label--diff congruence}, a measure of whether assigned labels correspond to the technical content of PR changes.

\paragraph{Code Review Dynamics and Collaboration Outcomes}
Research on PRs evaluation has highlighted factors influencing review speed and participation. Yu et al.~\citep{yu2015} found that the number of comments is the strongest predictor of latency, followed by PR size and the presence of continuous integration. Zhao et al.~\citep{zhao2019} developed a recommender for PRs likely to be reviewed quickly and reported, via survey, that time-based recommendations are valued by developers. Khatoonabadi et al.~\citep{khatoonabadi2024} predicted first-response latency, showing that submission timing, commit count, and contributor experience are major predictors and generalize across projects. Ruangwan et al.~\citep{ruangwan2019} analysed reviewer participation and found that 16--66\% of patches had at least one invited reviewer who did not respond, with prior participation rate and authoring experience strongly associated with responsiveness.  
While these studies link technical and social factors to review outcomes, we test whether metadata quality—specifically label--diff congruence—shapes review speed and discussion patterns.

\paragraph{Contributor Experience and Activity Patterns}
Contributor experience influences engagement with project processes~\citep{11025661}. Steinmacher et al.~\citep{steinmacher2015} identified 58 barriers faced by newcomers, including 13 social ones, and found that less-experienced contributors report more barriers regardless of project characteristics. Fronchetti et al.~\citep{fronchetti2023} studied CONTRIBUTING files from 2{,}274 projects and found that 52\% lacked guidance on at least three of six barrier categories, with task selection and community contact often missing. Pinto et al.~\citep{pinto2016} reported that one-time contributors are often motivated by bug fixing but that 47\% encounter barriers that hinder continued participation. Santos et al.~\cite{10.1145/3696630.3727243} found that intrinsic motivation predicts contribution interest more than perceived barriers, and that newcomers frequently misidentify the skills required for OSS tasks~\cite{9825764} (f-measure=0.37), underscoring the role of accurate issue labeling in task self-selection.
Building on this evidence, we test whether label--diff congruence differs systematically across contributor activity tiers, assessing whether coordination signal quality varies with experience.

\paragraph{Human and Social Aspects of Code Review}
Studies highlight the social dimension of review. Bacchelli and Bird~\citep{bacchelli2013} showed that developers value review for knowledge transfer, team awareness, and shared ownership in addition to defect detection; Bosu et al.~\citep{7180075} further found that reviewer experience and change size significantly affect the proportion of useful review comments. McIntosh et al.~\citep{mcintosh2016} showed that review coverage, participation, and reviewer expertise each share a significant link with post-release software quality. Alami et al.~\citep{alami2025} examined intrinsic drivers such as integrity, reputation, and accountability, finding that responsibility shifts from individual to collective during review and depends on hierarchy. Ciancarini et al.~\citep{ciancarini2023} reported that 75\% of developers believe interpersonal relationships can hinder objectivity; Fatima et al.~\citep{8975708} systematically identified individual, social, and personnel factors---including trust, emotions, and team interaction---that shape the review process. 
Whereas these works emphasize social and motivational factors, our study considers how technical metadata quality interacts with review by analyzing whether label--diff congruence is associated with discussion intensity and participant diversity.

%% file: dataset.tex
\section{Dataset}
\label{sec:dataset}
We analyse the public GitHub repository kubernetes/kubernetes using a dataset made available by Destefanis et al.~\cite{destefanis2026mining}. We focus on Kubernetes because it provides both the scale (26,000 contributors, 83,368 PRs over 11 years) and structure (175 distinct labels tied to Special Interest Groups) necessary to study label--code alignment and its coordination effects across diverse contributor populations. The formal labeling system and high activity volume enable robust stratification by contributor experience---a key requirement for testing whether alignment affects different populations differently.

The dataset integrates (i)~pull requests (PRs), (ii)~commits with file lists, and (iii)~issue and PR comments, with timestamps in UTC. PRs are the unit of analysis because they connect labels, code changes, and review discussions: each PR includes \texttt{area/*} labels declaring its scope, merged PRs provide diffs identifying modified files, and comments capture collaborative interaction. Author identifiers are anonymized and bots are identified in the dataset via a naming convention (\texttt{bot\_\#\#\#\#} pseudonyms)~\cite{destefanis2026mining}; we exclude bot-authored entries from all author- and discussion-level analyses---bot activity is known to influence OSS discussion dynamics~\cite{vaccargiu2026boatse}---retaining bot-authored merge commits solely to ensure diff coverage. Bot counts are reported in Table~\ref{tab:dataset_overview}; the commit snapshot extends to 2025-07-01 solely to verify file-list availability for merge commits.

%% file: methodology.tex
\section{Methodology}
\label{sec:method}

This is an exploratory case study investigating how metadata quality affects coordination among contributors with different levels of project familiarity—a fundamentally human question about who can navigate project structure, attract reviewer attention, and coordinate effectively during distributed collaboration. \textbf{Our approach combines large-scale measurement with behavioral validation.} We measure label--code alignment across 18,020 pull requests to establish prevalence and patterns (RQ1), then test associations with collaboration outcomes while stratifying by contributor experience to examine heterogeneity (RQ2). 

Behavioral validation comes from two sources: (i)~text mining of 1M+ comments shows developers actively discuss coordination (40\% of PRs) and modify labels (9\% of PRs), and (ii)~systematic patterns (e.g., modifications peak at partial alignment, strongly predicted by initial congruence, odds ratio (OR) =0.740) indicate developers monitor and maintain alignment as part of review practice. \textbf{While we do not conduct interviews, we document observable coordination behaviors}—labeling, correcting, routing—that reveal how contributors use metadata signals in practice. The scale of analysis (18K PRs, 9 years) provides statistical power to detect differential effects across contributor populations, though we interpret findings as associations that illuminate coordination patterns rather than causal claims about individual behavior.

\begin{figure*}[t]
\centering
\includegraphics[width=\textwidth]{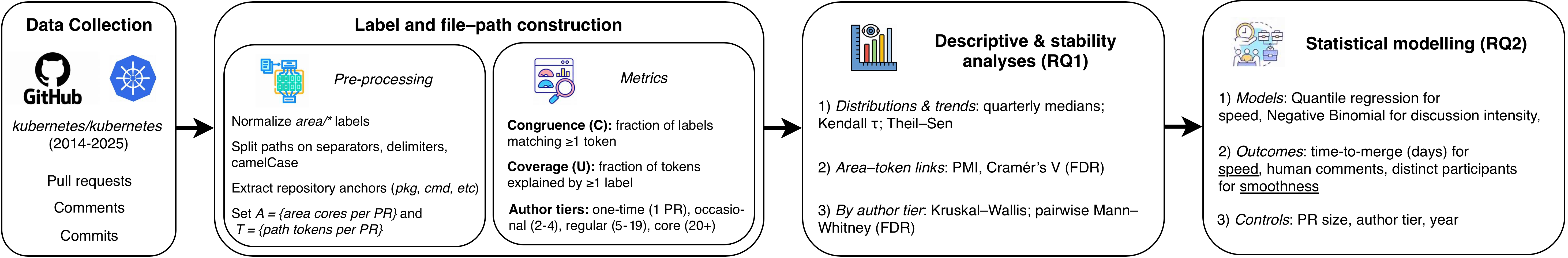}
\caption{Overview of the methodology. Data from the Kubernetes repository is processed to construct label--diff congruence metrics, which are then analyzed to assess prevalence, stability, and interpretability (RQ1), and their relationship with collaboration outcomes (RQ2).}
\label{fig:methodology}
\end{figure*}

The analysis proceeds in four steps (Figure~\ref{fig:methodology}): (i)~data collection and cleaning for PRs, merge commits, and comments from the Kubernetes repository; (ii)~construction of label and file-path features to operationalize congruence; (iii)~descriptive and temporal analyses to assess prevalence, stability, and interpretability, including behavioral validation via comment text mining; and (iv)~statistical modeling of collaboration outcomes—time-to-merge via quantile regression, discussion intensity via negative binomial models—with stratification by contributor tier to test whether alignment affects populations in distinct ways. \textbf{Throughout, we prioritize transparency: all preprocessing choices are documented, integrity checks are explicit, and model diagnostics are reported.}

\subsection{Data collection and feature construction}

The study combines three sources: (i)~PR metadata (identifier, author, timestamps, labels, merge SHA), (ii)~merge commit records with file paths, and (iii)~PR comments with author and timestamp. We retain PRs meeting four criteria: (1)~merge SHA present, (2)~SHA joins to non-empty file list, (3)~at least one \texttt{area/*} label (also matching \texttt{area:} and \texttt{area-} variants), and (4)~human authorship (bots excluded). Applying all four criteria yields 18,020 PRs (18,025 pass criteria 1--3; five bot-authored PRs are excluded by criterion 4). These criteria ensure we can measure alignment: (1) and (2) provide the file-level diff needed to tokenize changes, (3) ensures contributors explicitly signaled scope through area labels, and (4) restricts analysis to human contributions. PRs without area labels either rely on other coordination mechanisms (assignees, SIG mentions) or lack explicit scope signals, making label--code alignment undefined. This filtering yields a sample where alignment is both measurable and theoretically meaningful---cases where contributors chose to use labels to signal scope. Author IDs are SHA256-hashed for anonymity while preserving linkage.

\paragraph{Label normalization and path tokenization.}
We extracted labels matching \verb|^area[/:\-](.+)$|, lowercased and trimmed the core, and standardized to \texttt{area/<core>}. Multiple area labels yield the distinct set $A$. For file paths, we split on path separators into segments. Token set $T$ always includes the first segment and the compound \texttt{first/second}. For any segment matching a root anchor (\texttt{pkg}, \texttt{cmd}, \texttt{staging}, \texttt{test}, \texttt{vendor}, \texttt{build}, \texttt{hack}, \texttt{api}, \texttt{cluster}, \texttt{config}, \texttt{contrib}, \texttt{docs}) anywhere in the path, we add that anchor and the compound \texttt{anchor/next}. All tokens are lowercased and deduplicated.

\paragraph{Metrics.} We calculated the following metrics:

\noindent\textbf{Congruence} $C$ is the fraction of area cores in at least one token:
\[
C = \frac{1}{|A|}\sum_{a\in A}\mathbf{1}[\exists t\in T: a \subseteq t].
\]
\textbf{Coverage} $U$ is the fraction of tokens explained by at least one area:
\[
U = \frac{1}{|T|}\sum_{t\in T}\mathbf{1}[\exists a\in A: a \subseteq t].
\]

Authors are grouped by lifetime PR count: one-time (1), occasional (2--4), regular (5--19), core ($\geq$20), used as factors in RQ1 and controls in RQ2.

\textit{Worked example.} PR with labels \texttt{area/kubelet}, \texttt{area/volume}; cores $A = \{\text{kubelet}, \text{volume}\}$. Files: \texttt{pkg/volume/csi/csi\_plugin.go}, \texttt{pkg/kubelet/volumemanager/volume\_manager.go}, \texttt{test/e2e/} \\ \texttt{storage/drivers.go}. Tokens $T = \{\texttt{pkg}, \texttt{pkg/volume}, \texttt{pkg/kubelet}, \\ \texttt{test}, \texttt{test/e2e}\}$. Both cores appear in $T$ (\texttt{volume} $\subset$ \texttt{pkg/volume}, \texttt{kubelet} $\subset$ \texttt{pkg/kubelet}), so $C = 2/2 = 1.0$.
Four of five tokens match a label core, so $U = 4/5 = 0.8$. \textbf{High congruence with lower coverage reflects coarse-grained label signals.}

\subsection{Descriptive and stability analyses (RQ1)}

\paragraph{RQ1—Prevalence and distribution.}
We assessed congruence distribution via medians, IQRs, and mass at 0 and 1. For prevalence at thresholds $\tau \in \{0, 0.25, 0.50, 0.75, 1\}$, we computed proportions $C \geq \tau$ with 95\% Wilson intervals:
\[
L_{\text{W}} = \frac{\hat p + \frac{z^2}{2n} - z\sqrt{\frac{\hat p(1-\hat p)}{n} + \frac{z^2}{4n^2}}}{1+\frac{z^2}{n}}, \quad
U_{\text{W}} = \frac{\hat p + \frac{z^2}{2n} + z\sqrt{\frac{\hat p(1-\hat p)}{n} + \frac{z^2}{4n^2}}}{1+\frac{z^2}{n}}
\]
with $z=1.96$. Wilson intervals remain accurate for extreme proportions (our sample: 47\% at $C=1$), unlike symmetric normal approximations that can extend beyond [0,1].

Wilson intervals are preferred over normal approximations because they remain accurate even when proportions are extreme (close to 0 or 1) and sample sizes are finite \cite{Wilson01061927}. In our case, 46.6\% of PRs have perfect congruence ($C=1$) and 35.9\% have zero congruence ($C=0$), making the distribution strongly bimodal with substantial mass at the boundaries. Symmetric intervals centered on such extreme proportions can produce bounds extending beyond [0,1], which Wilson's method prevents through asymmetric adjustment.

\paragraph{RQ1—Temporal stability.}
To test whether congruence was consistent over time, we aggregated PRs by calendar quarter and computed quarterly medians of $C$. We then estimated trend direction and magnitude using Kendall's $\tau_b$ between quarter index and medians \cite{kendall1938new}, and a Theil–Sen slope (robust median-of-slopes estimator) \cite{Sen01121968}. The fitted robust line was overlaid on the quarterly series. A non-significant slope supported treating congruence as stable across the project history, whereas a significant slope indicated systematic drift to be considered in subsequent analyses.

\paragraph{RQ1—Interpretability via area--token associations.}
We tested whether area cores aligned with technical domains by analyzing PR-level pairs $(a,t)$. We applied the following thresholds to focus on substantive patterns: area support $\geq 100$ PRs (appears in $>0.5\%$ of sample), token support $\geq 100$ PRs (likewise $>0.5\%$), co-occurrence $\geq 20$ (sufficient observations for stable estimates), PMI $\geq 1$ (doubling of expected frequency) \cite{church1990word}, and Cramér's $V \geq 0.10$ (non-trivial effect size by conventional standards) \cite{cramer1999mathematical}. After filtering, we quantified associations via:
\[
\mathrm{PMI}(a,t) = \log\frac{p(a,t)}{p(a)p(t)}, \quad
V = \sqrt{\frac{\chi^2}{n \cdot \min(r-1,c-1)}}
\]
PMI measures co-occurrence strength (PMI=3.59 means $2^{3.59} \approx 12\times$ more than chance); $V$ measures effect size ($V \geq 0.10$ = non-trivial). Multiple testing controlled via Benjamini--Hochberg ($\alpha=0.05$) \cite{benjamini1995}. Significant pairs validate that congruence captures meaningful alignment, not accidental matches.

\paragraph{RQ1—Behavioral validation.}
To validate that developers attend to alignment, we mined comment text for coordination signals and label modifications. We inferred label modifications from comment text because this approach captures changes that maintainers \emph{communicated} to contributors (e.g., \texttt{/label area/X} commands, bot confirmations, explicit instructions like ``please add area/Y''), which reflects the coordination practices visible to participants. While this likely underestimates silent edits, it captures the subset of modifications that are socially marked—the ones that shape contributors' understanding of labeling norms and expectations. We searched 1M+ human comments for patterns indicating: (i)~label discussion (e.g., ``wrong label'', ``add area/X''), (ii)~routing (\texttt{@mentions}, \texttt{/sig} commands), (iii)~scope clarifications (``out of scope''), and (iv)~label changes (bot notifications, \texttt{/label} commands, explicit instructions). Using 16 regular expressions, we flagged PRs with coordination discussion and label modifications. Logistic regressions \cite{hosmer2013applied} tested whether congruence predicted discussion presence (controlling for PR size) and modification likelihood, with standardized predictors and bootstrap CIs (1,000 iterations). These analyses address construct validity by examining whether low congruence associates with behaviors consistent with developers noticing and correcting misalignment.

\paragraph{RQ1—High-coverage sensitivity.}
Availability of file-level diffs varied over time, which could bias prevalence and stability estimates. To check robustness, we computed the quarterly merge–diff joinability ratio (the share of merged PRs with a non-empty file list) and re-estimated all RQ1 summaries on a high-coverage subset defined by a minimum ratio threshold. Agreement between full-sample and high-coverage results indicated that conclusions did not hinge on periods with limited diff availability; discrepancies flagged potential selection bias, discussed in Section~\ref{sec:threats}.

\paragraph{RQ1—Label modifications during review.}
To assess whether contributors actively monitor and correct label–code alignment, we examined whether labels are modified during the review process. Because GitHub timeline events (which record exact label addition and removal timestamps) were not available in our dataset, we inferred label modifications from comment text. We searched for patterns indicating that a label had been changed, including: (i)~bot notifications (e.g., ``added label,'' ``removed label''), (ii)~Kubernetes-specific label commands (\texttt{/label area/X}, \texttt{/remove-area Y}), and (iii)~explicit human instructions (e.g., ``please add label,'' ``change label to''). We constructed 16 regular expressions covering these categories and applied them to all human-authored comments. At the PR level, we created a binary indicator \textit{label\_modified} equal to 1 if any comment contained a modification signal and 0 otherwise.

To test whether initial label–code misalignment prompts corrective action, we estimated a logistic regression predicting \textit{label\_modified} from initial congruence, controlling for PR size (log-transformed file count). Predictors were standardised to facilitate interpretation of odds ratios, and 95\% confidence intervals were computed via bootstrap resampling (1,000 iterations). This analysis provides behavioral evidence of whether developers respond to label–code misalignment by adjusting labels during review, thereby validating that congruence reflects a coordination property that contributors actively maintain rather than passively accept.

\subsection{Statistical modelling (RQ2)}
We tested whether label–diff congruence was associated with two collaboration outcomes: \emph{speed} (time to merge) and \emph{discussion intensity} (comments and distinct participants). The analysis considered (i) time to merge in days, (ii) human comment count, and (iii) distinct human participants.

\paragraph{RQ2—Time to merge.}
Time-to-merge was skewed with no censoring (100\% merged). We used quantile regression \cite{koenker1978} because: (i)~it makes no distributional assumptions about residuals, (ii)~it tests effects at multiple points in the distribution (fast, typical, slow merges), revealing whether congruence affects routine PRs differently than complex ones, and (iii)~it is robust to extreme outliers (e.g., the maximum time-to-merge of 1,461 days would not unduly influence median estimates). For $\tau \in \{0.25, 0.50, 0.75\}$ we estimated:
\[
Q_Y(\tau|X) = \beta_0(\tau) + \beta_1(\tau)C + \beta_2(\tau)\log(1+\text{files}) + \boldsymbol{\gamma}(\tau)^\top \mathbf{T} + \boldsymbol{\delta}(\tau)^\top \mathbf{Y}
\]
where $C$ is congruence, files controls PR size, $\mathbf{T}$ encodes tier, $\mathbf{Y}$ encodes year. Quantile regression tests effects across the distribution: at $\tau=0.50$, does congruence affect median merge time? At $\tau=0.75$, does it affect slower PRs? This detects heterogeneity (e.g., congruence might help routine PRs but not complex ones). We reported robust SEs and FDR-corrected $p$-values across $\tau$.

\paragraph{RQ2—Discussion intensity.}
Comment and participant counts were overdispersed (variance/mean: 49.94 for comments, 1.32 for participants) and duration-dependent. We used Negative Binomial models with log link and exposure offset to estimate per-day rates while accounting for how long PRs stayed open \cite{cameron2013regression}. The offset term $\log(\text{duration days})$ adjusts for exposure: PRs open for 10 days naturally accumulate more comments than those open for 2 days, so we model daily rates rather than raw counts. For outcome $Y$ (comments or distinct participants):
\[
\log \mathbb{E}[Y|X] = \alpha + \theta C + \eta \log(1+\text{files}) + \boldsymbol{\phi}^\top \mathbf{T} + \boldsymbol{\psi}^\top \mathbf{Y} + \log(\text{duration days})
\]

The resulting IRRs quantify proportional changes in daily rates: IRR=0.957 means 4.3\% fewer comments \textit{per day} with higher congruence, regardless of duration. Negative Binomial (vs. Poisson) handles overdispersion via a dispersion parameter, yielding accurate SEs.

\paragraph{RQ2—Stratification by contributor tier.}
Since RQ1 showed core contributors tolerate lower congruence, we tested whether effects differ by experience. We estimated separate models per tier to allow effect directions to reverse across groups (one-time: IRR$>1$; core: IRR$<1$) and to let all coefficients vary by experience level, not just congruence. We re-estimated NB models separately per tier (one-time, occasional, regular, core), excluding tier dummies and retaining only congruence, size, and duration.

\paragraph{RQ2—Robustness and sensitivity.}
File-diff availability varied over time. To test robustness, we re-estimated all RQ2 models on a high-coverage subset defined by a minimum quarterly merge–diff joinability ratio. Consistent results across full and high-coverage samples indicated that conclusions did not depend on coverage fluctuations.

%% file: rq1.tex
\section{Prevalence, stability, and interpretability of label--diff congruence}
\label{sec:rq1}

\paragraph{Analysis sample and measurement context.}
Our analysis focuses on PRs where both labels and code changes are observable, allowing us to assess whether declared scope (via area labels) aligns with actual modifications. From 83,368 PRs in the Kubernetes repository, we identified 18,020 (21.6\%) that met four criteria: (i)~a merge commit with file-level diffs available, (ii)~at least one \texttt{area/*} label assigned, (iii)~human authorship (bots excluded), and (iv)~successful join between merge SHA and non-empty file list. This represents contributions where contributors explicitly signaled scope through labels and where we can verify that signal against the files they modified.

The filtering reflects practical constraints in the data: 99.0\% of PRs have merge commit identifiers, but only 49.7\% could be joined to complete file lists (coverage varied over time due to GitHub API limitations). Among all PRs, 37.9\% use \texttt{area/*} labels, indicating that this labeling practice, while common, is not universal. Our analysis set thus represents PRs where contributors chose to use area labels and where we have technical visibility into their changes—a context where label--code alignment can be measured and interpreted.

\paragraph{Label--code alignment is common but concentrated at extremes.}
Within the analysis set ($n=18{,}020$), the median label--diff congruence is 0.50, with an interquartile range spanning [0, 1]. The distribution is strongly bimodal: 46.6\% of PRs achieve perfect alignment ($C=1$), where every assigned label matches at least one modified file, while 35.9\% show zero alignment ($C=0$), where no labels correspond to changes. At a threshold of 0.75, representing substantial but not necessarily perfect alignment, 46.7\% of PRs meet or exceed this level.

This pattern is largely shaped by labeling practice: the median number of \texttt{area/*} labels per PR is just 1 (25th--75th percentile: 1--2). When contributors assign a single label, it either fully captures the scope of their changes or misses entirely—there is no middle ground. The prevalence of perfect alignment (nearly half of all PRs) indicates that \textbf{many contributors successfully signal their work's scope through labels}. However, the substantial minority with zero alignment suggests cases where labels and changes diverge, potentially creating coordination friction as reviewers or maintainers encounter unexpected scope.

As a complementary perspective, we measured coverage: the proportion of file-path tokens that are explained by at least one assigned label. The median coverage is 0.3077, meaning that labels typically account for fewer than one-third of the structural information present in modified file paths. This low coverage does not necessarily indicate poor labeling—it may reflect that \textbf{labels are intended as coarse-grained scope signals} rather than exhaustive file manifests. A label like \texttt{area/kubelet} correctly signals that changes affect the kubelet subsystem without enumerating every modified file within that area. This distinction matters for interpreting alignment: high congruence with low coverage means labels are accurate but partial, a design choice rather than a deficiency.

The median congruence is 0.50 (IQR [0, 1]), with 46.6\% achieving perfect alignment ($C=1$), 35.9\% showing zero alignment ($C=0$), and 46.7\% exceeding 0.75. The median number of area labels per PR is 1 (P25--P75: 1--2), explaining the discrete mass at extremes. Median coverage is 0.308, indicating labels provide coarse-grained scope signals.

\begin{figure}[h]
  \centering
  \includegraphics[width=\linewidth]{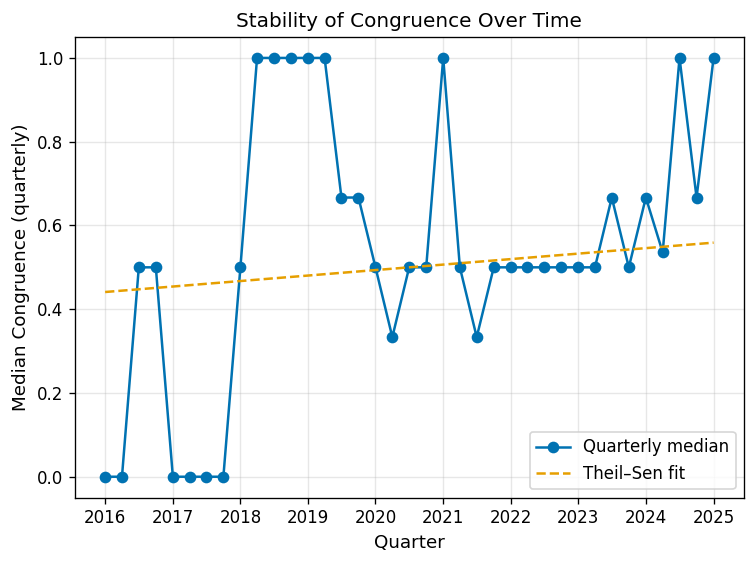}
  \caption{Quarterly median congruence with Theil--Sen robust fit (2016 Q1 to 2025 Q1). The modest upward trend indicates stable labeling practices over nine years of project growth.}
  \label{fig:rq1-trend}
\end{figure}

\paragraph{Labeling practices are stable over the project's history.}
To test whether label--code alignment changed systematically as the project matured, we aggregated PRs by creation quarter from 2016 Q1 to 2025 Q1 (37 quarters, spanning nine years). Per-quarter sample sizes ranged from 33 to 957 PRs (median: 488), providing sufficient data across the project's timeline. Quarterly median congruence fluctuated modestly but showed a small upward trend (Figure~\ref{fig:rq1-trend}): Kendall's $\tau_b = 0.298$ ($p=0.0178$) indicates a positive monotonic association, and the Theil--Sen robust slope estimate is 0.0033 per quarter (95\% CI: [0, 0.0192]).

\textbf{This suggests that label--code alignment is a persistent feature of Kubernetes coordination rather than an artifact of early development or recent policy changes.} The small magnitude of the trend (roughly 0.03 increase in median congruence per year) indicates that labeling practices have remained relatively consistent despite the project's growth from a few hundred to tens of thousands of contributors. For practical purposes, congruence can be treated as stable, meaning the patterns we document reflect enduring coordination practices rather than transient phenomena.

\vspace{-10pt}

\paragraph{High-congruence labels correspond to recognizable technical domains.}
To evaluate whether area labels align with coherent subsystems—meaning labels like \texttt{area/storage} consistently map to storage-related files—we analyzed associations between label cores and file-path tokens at the PR level. We applied objective thresholds to avoid spurious patterns: area support $\geq 100$ PRs, token support $\geq 100$ PRs, co-occurrence $\geq 20$, pointwise mutual information (PMI) $\geq 1$ bit, and Cramér's $V \geq 0.10$. After multiple testing correction using Benjamini--Hochberg FDR ($\alpha=0.05$), six area--token pairs survived all filters (Table~\ref{tab:rq1-areas}).

These associations confirm that certain labels systematically correspond to identifiable parts of the codebase. For example, \texttt{provider/} \texttt{openstack} co-occurs strongly with \texttt{pkg/cloudprovider} (PMI = 3.59, $V = 0.19$), and \texttt{conformance} aligns with \texttt{test/conformance} (PMI = 1.69, $V = 0.18$). \textbf{This indicates that high congruence is not an accident of string matching but reflects meaningful alignment between declared scope and technical reality.} Contributors who assign \texttt{area/conformance} to PRs modifying conformance tests are signaling accurately, and reviewers familiar with that subsystem can use the label as a reliable routing cue. The interpretability of these associations validates that congruence captures a coordination property developers recognize and use, rather than an artifact of our tokenization choices.

\begin{table}[t]
\centering
\small
\setlength{\tabcolsep}{3pt}
\caption{Top area--token associations after objective filters and FDR correction. High PMI and Cram\'er's $V$ indicate meaningful alignment between labels and file structure.}
\label{tab:rq1-areas}
\begin{tabular}{l l r r r r r}
\toprule
\textbf{Area} & \textbf{Token} & $n_{\text{area}}$ & $n_{\text{token}}$ & $n_{\cap}$ & \textbf{PMI} & $V$ \\
\midrule
provider/openstack & pkg/cloudprovider & 271 & 353  & 64  & 3.592 & .191 \\
node-e2e           & test/e2e\_node    & 432 & 1832 & 143 & 1.703 & .118 \\
conformance        & test/conformance  & 2475 & 716 & 318 & 1.694 & .181 \\
release-eng        & CHANGELOG         & 4906 & 174  & 136 & 1.522 & .112 \\
ipvs               & pkg/proxy         & 1960 & 986  & 280 & 1.385 & .135 \\
provider/gcp       & cluster/gce       & 3907 & 1137 & 506 & 1.038 & .143 \\
\bottomrule
\end{tabular}
\end{table}

\begin{table}[t]
\centering
\small
\setlength{\tabcolsep}{4pt}
\caption{Pairwise tier comparisons (FDR-corrected). Small positive 
effect sizes indicate core contributors have modestly higher congruence.}
\label{tab:rq1-tiersstats}
\begin{tabular}{l r r}
\toprule
\textbf{Test} & \textbf{p (FDR)} & \textbf{Effect} \\
\midrule
Kruskal--Wallis (all tiers) & $1.84{\times}10^{-12}$ & $H=57.683$ \\
Core vs.\ regular           & $7.72{\times}10^{-8}$  & $\delta=+0.069$ \\
Core vs.\ occasional        & $3.83{\times}10^{-6}$  & $\delta=+0.095$ \\
Core vs.\ one-time          & $7.70{\times}10^{-4}$  & $\delta=+0.084$ \\
\bottomrule
\end{tabular}
\end{table}

\paragraph{Label--code alignment varies with contributor experience.}
We grouped contributors by their lifetime PR count in the dataset: one-time (1 PR), occasional (2--4 PRs), regular (5--19 PRs), and core ($\geq$20 PRs). The analysis set comprises 14,687 PRs from core contributors, 2,132 from regular, 736 from occasional, and 465 from one-time contributors. Rather than comparing full distributions (which are heavily bimodal), we examined the proportion of PRs at the distribution's extremes: perfect alignment ($C=1$) and zero alignment ($C=0$), with 95\% Wilson confidence intervals.

\textbf{Core contributors achieve perfect alignment more frequently and zero alignment less frequently than other tiers.} Specifically, 47.5\% of core PRs have $C=1$ compared to 44.5\% for one-time contributors, and 34.3\% of core PRs have $C=0$ compared to 46.5\% for one-time contributors. A Kruskal--Wallis test \cite{Kruskal01121952} confirms differences across tiers ($H=57.683$, $p=1.84 \times 10^{-12}$), and pairwise Mann--Whitney tests \cite{mann1947} (FDR-corrected) show small but consistent shifts favoring core contributors (Cliff's $\delta \approx 0.069$--0.095, equivalent to $A_{12} \approx 0.534$--0.547; Table~\ref{tab:rq1-tiersstats}).

\textbf{These differences concentrate at the tails, suggesting distinct coordination challenges.} Experienced contributors more often achieve full alignment, possibly because they understand subsystem boundaries and label conventions through repeated participation. Less experienced contributors more often face complete mismatches, which may reflect unfamiliarity with how labels map to code structure or uncertainty about which area to declare. This pattern foreshadows the stratified analysis in RQ2, where we examine whether alignment matters differently for contributors with varying levels of project knowledge.

\paragraph{Congruence declines with PR size, as expected.}
As context for subsequent analyses, we note that congruence declines as PRs modify more files. For small PRs ($\leq 2$ files, $n=7{,}559$) and medium PRs (3--5 files, $n=5{,}739$), the median congruence is 1.0; for large PRs ($\geq 6$ files, $n=4{,}727$), the median is 0.5. A Kruskal--Wallis test confirms strong differences ($H=360.742$, $p=4.63\times 10^{-79}$). \textbf{This is unsurprising: as changes span more files, they are more likely to cross subsystem boundaries, making single-label descriptions inadequate.} We control for PR size in subsequent regression models to isolate the effect of alignment from the confounding influence of change scope.

\vspace{-5pt}

\paragraph{Developers actively discuss coordination, though explicit label complaints are rare.}
Among the 18,020 PRs, 7,263 (40.3\%) contained coordination-related comments, most commonly \texttt{@mentions for routing} (28,854 instances) and \texttt{/sig} commands (9,333 instances). Explicit label complaints were rare (518 instances). Logistic regression found no association between congruence and coordination discussion (OR = 1.017, $p > 0.05$), suggesting developers route work using social mechanisms regardless of label accuracy.

\vspace{-5pt}

\paragraph{Labels are actively refined during review, especially when partially aligned.}
While explicit complaints are rare, \textbf{label modifications are common}: 1,654 PRs (9.2\%) showed evidence of labels being changed during review, inferred from comment text patterns such as bot notifications (``added label area/X''), \texttt{/label} commands, or explicit instructions (``please add area/Y''). The proportion of PRs with modifications varied systematically by initial congruence: 11.1\% at $C=0$, 14.1\% at $C \in (0, 0.5]$, \textbf{21.7\%} at $C \in (0.5, 1)$, and 5.6\% at $C=1$. The elevated rate in the middle range is striking: \textbf{maintainers are most attentive to PRs with partial alignment}, refining labels when they are ``close but not quite right'' rather than leaving obvious mismatches or perfect matches unchanged.

Logistic regression confirmed that initial congruence strongly predicts label stability: OR = 0.740 (95\% CI [0.705, 0.778], $p < 0.001$). Higher initial congruence is associated with 26\% lower odds of modification. \textbf{This systematic correction—without extended debate—provides direct behavioral evidence that developers treat alignment as worth maintaining, validating that congruence captures a coordination property contributors actively monitor rather than an abstract measurement artifact.}


\begin{tcolorbox}[right=0.1cm,left=0.1cm,top=0.1cm,bottom=0.1cm]
\textbf{Answer to RQ1:}  
Label--diff congruence is \textbf{prevalent} (median 0.50; 47\% of PRs at perfect alignment, 36\% at zero), \textbf{stable} over nine years of project history (small upward trend), and \textbf{interpretable} (surviving area--token associations map to recognizable subsystems after objective filtering and FDR correction). \textbf{Core contributors achieve higher alignment than less experienced contributors}, and congruence declines with PR size, indicating that both experience and change scope shape labeling accuracy. \textbf{Behavioral validation} shows that developers actively maintain alignment: 9.2\% of PRs have labels modified during review, with corrections concentrated in the partial-alignment range, providing direct evidence that contributors monitor and refine label--code correspondence as part of coordination practice.
\end{tcolorbox}

%% file: rq2.tex
\vspace{-8pt}
\section{Congruence, speed, and smoothness on Kubernetes PRs}
\label{sec:rq2}

\paragraph{Measuring collaboration through review patterns.}
We examine whether label--code alignment relates to how contributors collaborate during review. Our analysis focuses on 18,020 merged PRs—the same set used in RQ1—and measures two dimensions of collaboration. \textbf{Speed} captures when decisions are made: how long from PR creation to merge. \textbf{Smoothness} captures the character of interaction: how much discussion occurs (human comment volume) and how many people participate (distinct human reviewers). These dimensions are conceptually distinct: a PR might merge quickly but involve extensive debate, or merge slowly with minimal discussion. All PRs in our analysis sample merged (100\%), so we model time-to-merge conditional on eventual acceptance rather than studying rejection.

The distributions of collaboration outcomes reflect Kubernetes' scale: time-to-merge has median 5.1 days (mean 24.2, SD 57.9) with substantial right skew (range 0--1,461 days); human comments have median 9 (mean 17.7, SD 29.8) with strong overdispersion (variance/mean = 49.94); distinct participants have median 3 (mean 3.8, SD 2.2), also overdispersed (variance/mean = 1.32). Distributions exhibit heavy right skew (durations) and overdispersion (comments, participants), motivating our choice of quantile regression and negative binomial models.

\paragraph{Label--code alignment does not predict merge speed.}
Quantile regression at $\tau \in \{0.25, 0.50, 0.75\}$ showed no significant relationship between congruence and time-to-merge. At the median, the coefficient is $-0.1241$ days (SE = 0.1730, $p=0.4730$; Table~\ref{tab:rq2_quantile}). Benjamini--Hochberg adjustment yields $p=1.0$ for all quantiles. \textbf{Predicted median time-to-merge remains essentially constant across the congruence range (approximately 5.1 days), indicating label--code alignment does not influence when decisions are made.} This suggests merge speed is driven by other factors—reviewer availability, change complexity, testing requirements—rather than label accuracy.

\begin{table}[h]
\centering
\small
\setlength{\tabcolsep}{3pt}
\caption{Quantile regression: congruence effect on time-to-merge (days). No significant relationship at any quantile after FDR correction.}
\label{tab:rq2_quantile}
\begin{tabular}{lrrr}
\toprule
\textbf{Quantile ($\tau$)} & \textbf{$\beta$} & \textbf{SE} & \textbf{$p$} \\
\midrule
0.25 & 0.0212 & 0.0637 & 0.7394 \\
0.50 & $-0.1241$ & 0.1730 & 0.4729 \\
0.75 & $-0.6502$ & 0.7326 & 0.3748 \\
\bottomrule
\multicolumn{4}{l}{\footnotesize FDR-adjusted $p=1.0$ for all quantiles. Spearman $\rho=-0.017$, $p=0.0198$.} \\
\end{tabular}
\end{table}

\vspace{-6pt}
\paragraph{Higher congruence predicts quieter reviews: fewer comments per day.}
For human comments, the congruence coefficient is $-0.044$ (SE = 0.018, $z=-2.43$, $p=0.015$), yielding IRR = 0.957 (95\% CI [0.923, 0.992]). A one-unit increase in congruence is associated with 4.3\% fewer human comments (IRR = 0.957, $p$ = 0.015), or roughly 0.75 fewer comments relative to the sample mean of 17.7. This modest but statistically robust reduction suggests that misaligned labels may generate additional clarification---reviewers asking about scope, maintainers redirecting attention---while aligned labels reduce the need for meta-discussion.

\paragraph{Higher congruence predicts focused reviews: fewer distinct participants.}
For distinct human participants (Table~\ref{tab:rq2_nb_participants}), the congruence coefficient is $-0.156$ (SE = 0.020, $z=-7.91$, $p < 0.001$), corresponding to IRR = 0.856 (95\% CI [0.824, 0.890]). A one-unit increase in congruence is associated with 14.4\% fewer distinct participants (IRR = 0.856, $p$ < 0.001), corresponding to approximately 0.5 fewer participants for a PR at the sample mean of 3.8. When labels accurately signal scope, relevant reviewers identify themselves more easily and reviews remain focused; when labels misalign, PRs attract reviewers from multiple areas as contributors notice the mismatch and redirect attention. This suggests accurate labels help contributors find the right reviewers more efficiently.

\begin{table}[h]
\centering
\small
\setlength{\tabcolsep}{3pt}
\caption{Negative binomial model: congruence effect on distinct participants (per day). Higher congruence predicts substantially fewer participants.}
\label{tab:rq2_nb_participants}
\begin{tabular}{lrrrr}
\toprule
\textbf{Covariate} & \textbf{Coef.} & \textbf{SE} & \textbf{IRR} & \textbf{$p$} \\
\midrule
Congruence & $-0.156$ & 0.020 & 0.856 & $<0.001$ \\
log(files) & $-0.311$ & 0.015 & 0.856 & $<0.001$ \\
\multicolumn{5}{l}{\footnotesize Tier and year controls included; $n=18{,}020$.} \\
\bottomrule
\end{tabular}
\end{table}

\paragraph{Effects differ substantially by contributor experience.}
The pooled estimates above average across all contributors, but \textbf{coordination needs may differ for newcomers versus experienced developers}. To test this, we re-estimated the negative binomial models separately for each contributor tier: one-time (1 PR, $n=465$), occasional (2--4 PRs, $n=736$), regular (5--19 PRs, $n=2{,}132$), and core ($\geq$20 PRs, $n=14{,}687$). Because tier was the stratifying variable, tier indicators were excluded from these models; we estimated only congruence and PR size effects within each group.

\begin{figure*}[t]
\centering
\includegraphics[width=0.85\textwidth]{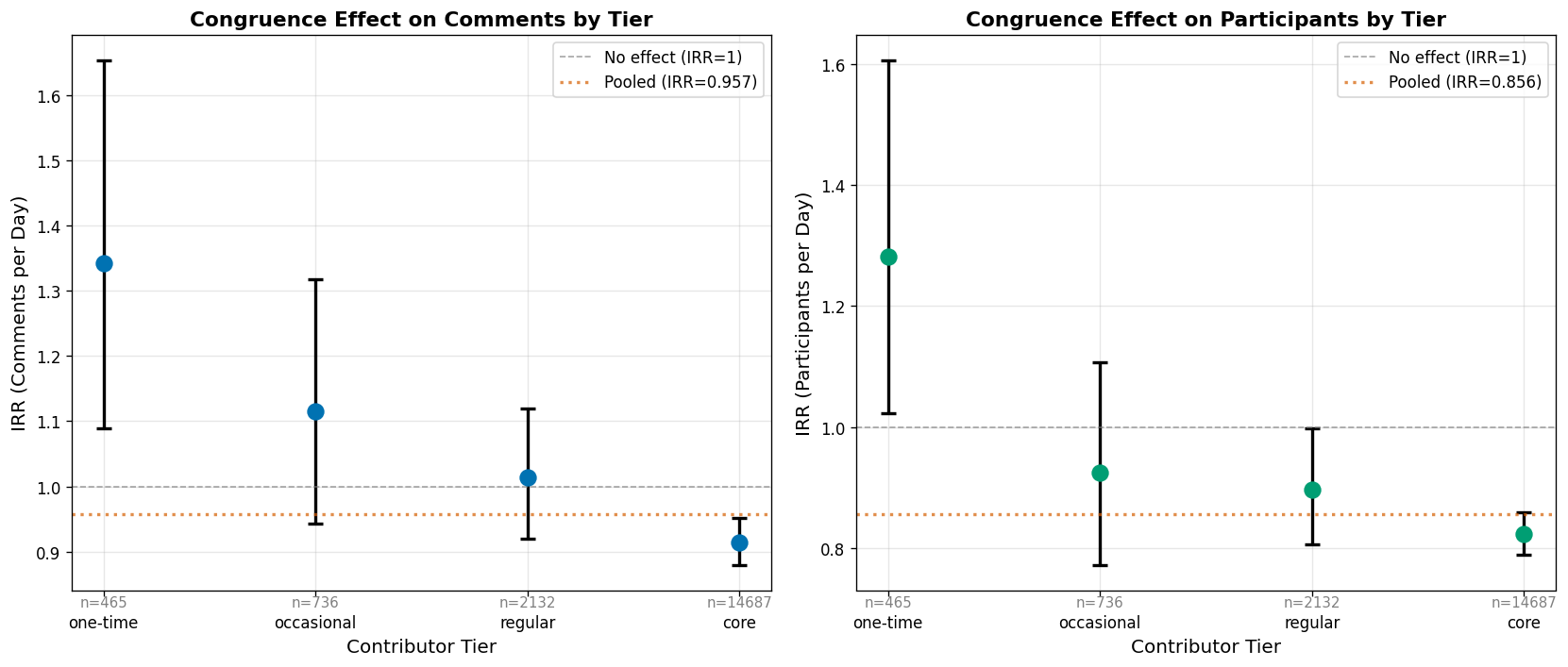}
\caption{Congruence effects by contributor tier (IRRs with 95\% CIs). Left: comments. Right: participants. Gray dashed line indicates no effect (IRR=1); red dotted line shows pooled estimate. Effects reverse direction for one-time versus core contributors.}
\label{fig:rq2_stratified}
\end{figure*}

\textbf{The pattern that emerges is striking} (Figure~\ref{fig:rq2_stratified}). Among \textbf{core contributors}, who represent 81\% of the sample, higher congruence predicts \emph{fewer} comments per day (IRR = 0.915, 95\% CI [0.880, 0.951], $p < 0.001$) and \emph{fewer} participants (IRR = 0.824, [0.790, 0.859], $p < 0.001$). This matches the pooled result and confirms the interpretation above: for experienced developers, accurate labels streamline coordination by reducing clarification needs and focusing participation.

However, among \textbf{one-time contributors}, the direction \emph{reverses}: higher congruence is associated with \emph{more} comments (IRR = 1.342, [1.089, 1.654], $p=0.006$) and \emph{more} participants (IRR = 1.282, [1.023, 1.606], $p=0.031$). For occasional and regular contributors, effects were intermediate or non-significant (Table~\ref{tab:rq2_stratified_summary}).

\paragraph{Interpreting heterogeneity: labels serve dual coordination functions.}
This heterogeneity reveals that label--code alignment affects coordination differently depending on contributor familiarity. \textbf{For core developers (81\% of the sample), higher congruence predicts quieter reviews} (IRR = 0.82 participants, 0.92 comments), consistent with reduced clarification needs when scope is clear. 

\textbf{For one-time contributors, the direction reverses} (IRR = 1.28 participants, 1.34 comments), suggesting accurate labels help newcomer contributions attract reviewer attention during triage. This dual-function interpretation—streamlining for experts, facilitating discovery for newcomers—reconciles the RQ1 paradox where core contributors showed lower congruence yet coordinated effectively. The pooled estimates reflect the dominant pattern among core contributors; stratified analysis reveals effects are not uniform across the contributor base.

\begin{table}[t!]
\centering
\footnotesize
\setlength{\tabcolsep}{3.5pt}
\caption{Congruence effects on discussion by tier. Opposite directions for newcomers vs.\ core developers indicate differential coordination functions.}
\label{tab:rq2_stratified_summary}
\begin{tabular}{@{}lrrrrr@{}}
\toprule
 & \multicolumn{2}{c}{\textbf{Comments}} & \multicolumn{2}{c}{\textbf{Participants}} & \\
\textbf{Tier} & \textbf{IRR [95\% CI]} & \textbf{$p$} & \textbf{IRR [95\% CI]} & \textbf{$p$} & $n$ \\
\midrule
One-time   & 1.342 [1.089, 1.654] & 0.006 & 1.282 [1.023, 1.606] & 0.031 & 465 \\
Occasional & 1.115 [0.944, 1.318] & 0.201 & 0.925 [0.772, 1.107] & 0.393 & 736 \\
Regular    & 1.014 [0.919, 1.119] & 0.780 & 0.897 [0.807, 0.998] & 0.045 & 2{,}132 \\
Core       & 0.915 [0.880, 0.951] & <0.001 & 0.824 [0.790, 0.859] & <0.001 & 14{,}687 \\
\bottomrule
\end{tabular}
\end{table}

\begin{tcolorbox}[right=0.1cm,left=0.1cm,top=0.1cm,bottom=0.1cm]
\textbf{Answer to RQ2:}  
Label--diff congruence does not affect collaboration \textbf{speed} (time-to-merge: $\beta=-0.1241$ days, $p=0.4660$; no significant effect at any quantile after FDR correction). In contrast, congruence affects \textbf{smoothness}: higher alignment predicts fewer human comments per day (IRR = 0.957, $p=0.015$) and fewer distinct participants (IRR = 0.856, $p<0.001$), controlling for PR size, author tier, and temporal trends. 

Among \textbf{core contributors} (81\% of the sample), higher congruence predicts fewer comments (IRR = 0.915) and participants (IRR = 0.824), consistent with reduced clarification needs when scope is clear. Among \textbf{one-time contributors}, the effect reverses: higher congruence predicts more comments (IRR = 1.342) and participants (IRR = 1.282), suggesting that accurate labels help newcomers attract reviewer attention during triage.

Overall, congruence influences \emph{how} collaboration proceeds rather than \emph{when} decisions are made, and labels serve different coordination functions depending on contributor experience.

\end{tcolorbox}

%% file: discussion.tex
\section{Discussion}
\label{sec:discussion}
We measured label--diff congruence—the alignment between PR labels and modified files—across 18,020 Kubernetes PRs. Congruence is common (47\% perfect alignment), stable over nine years, and corresponds to recognizable subsystems. Developers actively maintain it: 9\% of PRs have labels modified during review, with the likelihood strongly predicted by initial congruence (OR = 0.740). Congruence does not affect merge speed but shapes review: in pooled models, higher alignment is associated with fewer comments (IRR = 0.957) and fewer participants (IRR = 0.856). Crucially, effects differ by contributor tier: core developers experience quieter reviews with higher congruence (IRR = 0.82 participants), whereas one-time contributors see more engagement (IRR = 1.28 participants).

These stratified findings indicate that label--code alignment plays distinct roles. For experienced developers (81\% of the sample), accurate labels streamline coordination: scope is clear, reviewers assess relevance quickly, and participation is more focused (IRR = 0.82 participants, IRR = 0.92 comments). For newcomers, the pattern reverses: accurate labels make their contributions more visible during triage, drawing in more reviewers and discussion (IRR = 1.28 participants, IRR = 1.34 comments). In this case, additional comments represent engagement rather than friction. In a large project with thousands of contributors, being reviewed—even with more interaction—is preferable to being overlooked.

This dual role reconciles the RQ1 paradox: core contributors often show lower congruence yet coordinate effectively because they can rely on alternative mechanisms such as \texttt{@mentions}, SIG membership, or tacit knowledge~\cite{vaccargiu2026greenoss}. Newcomers lack these channels, making labels a critical entry point. The coexistence of formal labeling and informal routing reflects socio-technical congruence: coordination arises from multiple mechanisms whose importance depends on context~\cite{cataldo2008socio}; contributor role shapes not only coordination behaviour but also communication patterns more broadly~\cite{VACCARGIU2026108003, 11024360}.

Two analyses confirm that developers attend to alignment. First, 40\% of PRs contain coordination-related discussion (28,854 \texttt{@mentions}, 9,333 \texttt{/sig} commands), showing active routing even when labels are imperfect. Second, 9\% of PRs have labels modified during review, with corrections concentrated at partial alignment ($C \in (0.5,1)$). This pattern suggests developers quietly refine labels that are “close but not quite right,” treating alignment as worth maintaining without extended debate. Accurate labels thus work invisibly, while inaccurate ones are corrected.
Kubernetes practices appear optimized for experienced contributors, who can coordinate effectively even with lower congruence. For newcomers, however, accurate labels are more consequential: they provide visibility in triage and reduce barriers to participation~\cite{steinmacher2015}. This suggests that supporting diverse contributors requires attention to label quality. Possible interventions include: (i) onboarding-aware triage that checks label accuracy in newcomer PRs and provides explanatory corrections; (ii) tooling that flags mismatches at submission with newcomer-friendly prompts~\cite{SANTOS2025107568,10173918}; and (iii) clearer documentation linking labels to code structure. Coordination infrastructure is not one-size-fits-all: what works for experts may not serve newcomers equally. For projects with similar labeling conventions, analogous diagnostics may apply. Monitoring alignment could serve as a lightweight indicator of coordination health; triage support could reduce misrouted reviews; and inclusive labeling practices could help projects with high turnover treat labels as part of onboarding infrastructure. These considerations connect to a broader literature on OSS onboarding 
barriers~\cite{steinmacher2015,pinto2016,fronchetti2023}: labels are part of the project infrastructure newcomers must navigate, and improving their quality is a concrete, low-cost lever for inclusion. They may also extend to automated triage systems, where congruence offers a computable signal for flagging mismatches before human review. This study shows that routine metadata shapes collaboration in measurable ways. Labels affect participation, discussion dynamics, and newcomer visibility—effects modest in magnitude (e.g., 0.8 fewer comments, 0.5 fewer participants) but systematic across thousands of PRs. The findings indicate that coordination signals do not function uniformly: the same mechanism streamlines interaction among experienced contributors while helping newcomers gain visibility, showing how technical artifacts intertwine with social processes and how metadata quality influences the contributor experience as much as project organization.

%% file: threats.tex
\vspace{-6pt}
\section{Threats to Validity}
\label{sec:threats}

We structure threats following Wohlin et al.~\cite{wohlin2012}.

\textbf{\textit{Construct validity.}} Label--diff congruence relies on token matching between area labels (e.g., \texttt{area/storage}) and file-path segments (e.g., \texttt{pkg/storage/volume.go}). This captures naming alignment but may miss semantic matches when terminology differs (e.g., \texttt{area/api-machinery} vs. \texttt{pkg/registry/}). We mitigated this by normalizing labels, applying consistent tokenization, complementing congruence with coverage, and validating results through association tests (PMI, Cramér's $V$) and behavioral evidence (label corrections). Congruence therefore reflects literal correspondence rather than semantic fit. Coverage is low (median 0.31), but the prevalence of perfect congruence (47\%) and active label maintenance (9\% corrections, few complaints) suggest labels function as coarse-grained scope signals rather than exhaustive descriptions.\\
\noindent
\textbf{\textit{Internal validity.}} Confounding factors include bot activity, PR size, contributor experience, and temporal variation. We excluded bots from discussion metrics and controlled for file count, activity tier, and calendar year in regressions. Residual influences such as reviewer workload or concurrent events cannot be excluded. Stratified analysis (RQ2) showed heterogeneity: for newcomers, congruence may either attract review or result from post-hoc label refinement. As we cannot establish causality, results are reported as associations.\\
\noindent
\textbf{\textit{External validity.}} Findings are based on Kubernetes, a large project with structured governance and formal labels tied to Special Interest Groups. This makes it a best-case setting for studying coordination signals at scale, but also limits transferability: most GitHub projects use labels informally or not at all, and effects may be weaker or more variable where labeling is inconsistent. What we measure—congruence, coverage, and discussion intensity—can be computed in any project with labels and diffs, so the construct itself is portable. The mechanisms we identify (labels as scope signals, differential effects by experience, active maintenance) are likely to appear in other structured projects (e.g., Linux kernel, Chromium), but replication is needed to assess generality beyond highly governed settings~\cite{WIERINGA2015136}.\\
\noindent
\textbf{\textit{Data coverage.}} Only half of merged PRs could be linked to complete diffs (50.1\% joinability), with coverage lower in early years. We reported joinability transparently and repeated analyses on high-coverage subsets ($\geq$60\%), obtaining consistent results. This limitation mainly affects prevalence estimates, while associations (RQ2) remain robust.\\
\noindent
\textbf{\textit{Statistical conclusion validity.}}
We applied FDR correction, reported effect sizes alongside $p$-values, and used robust estimators (Theil--Sen, Negative Binomial). Effects are often modest in size (e.g., 0.8 fewer comments per PR), and stratified findings for newcomers are based on smaller samples, widening uncertainty. We therefore interpret subgroup effects as exploratory, while pooled estimates reflect the dominant core contributor population.

%% file: conclusion.tex
\section{Conclusion}
\label{sec:conclusion}

This case study introduced label--diff congruence, a measure of alignment between PR labels and modified files, and analyzed its role in coordination within Kubernetes. Across 18,020 PRs spanning nine years, congruence proved prevalent, stable, and actively maintained, with developers correcting labels during review in predictable ways. This confirms that contributors treat label–code correspondence as part of routine coordination.

Congruence does not influence merge speed but shapes how reviews unfold. Higher alignment is linked to fewer comments and participants overall, yet stratified analysis shows divergent effects: for core developers, it streamlines coordination by reducing discussion, while for newcomers it increases visibility and engagement. Labels therefore serve dual functions, supporting efficiency for experienced contributors and discovery for first-time participants.

These findings highlight the importance of metadata quality for inclusive collaboration. Monitoring congruence, flagging mismatches at submission, and providing clearer documentation can lower barriers for newcomers who rely on formal signals, while also reducing coordination overhead for experts. 

More broadly, the results demonstrate that routine metadata subtly shapes collaboration patterns. Labels may not determine merge outcomes, but they affect participation, discussion, and contributor experience. Small artifacts of process thus connect technical structure with social organization, influencing who can effectively engage in large-scale projects with structured labeling conventions.